\documentclass[11pt]{article}
\usepackage{amsfonts,amsmath,enumerate}

\usepackage{graphicx}

\def\R{{\mathbb R}}
\def\C{{\mathbb C}}

\begin{document}

\title{Faddeev eigenfunctions for two-dimensional Schr\"odinger operators via the
Moutard transformation}
\author{I.A. Taimanov
\thanks{Sobolev Institute of Mathematics, 630090 Novosibirsk, Russia;
e-mail: \texttt{taimanov@math.nsc.ru}} \and S.P. Tsarev
\thanks{Siberian Federal University, 79 Svobodny Prospect,
Krasnoyarsk 660041 Russia; e-mail: \texttt{sptsarev@mail.ru}.}}

\date{}

\maketitle

\abstract{We demonstrate how the Moutard transformation of
two-dimen\-sio\-nal Schr\"odinger operators acts on the Faddeev
eigenfunctions on the zero energy level and present some explicitly
computed examples of such eigenfunctions for smooth fast decaying
potentials of operators with non-trivial kernel and for nonstationary
potentials which correspond to blowing up solutions of the
Novikov--Veselov equation.}

\section{Introduction}

In \cite{TT} we used the Moutard transformation for constructing examples of
two-dimensional Schr\"odinger operators with fast decaying smooth potentials
and nontrivial kernels and of blowing up solutions of the Novikov--Veselov
equation.

Here we show how the Moutard transformation acts on
the Faddeev eigenfunctions corresponding to the zero energy level. In particular, we
construct some explicit examples which correspond to operators constructed in
\cite{TT}. It appears that such Faddeev eigenfunctions have very interesting
analytical properties and dynamics under the Novikov--Veselov flow.

The presented procedure also enables us to extend in an essentially new way
a list of two-dimensional Schr\"odinger operators for which
the Faddeev eigenfunctions are explicitly computed at least
for one energy level.

For multi-dimensional Schr\"odinger operator
$$
L = - \Delta + u
$$
the Faddeev eigenfunctions $\psi(x,k), k \in \C^n, x \in \R^n$, are formal solutions
to the equation
$$
L \psi = E \psi, \ \ \ E = \langle k,k \rangle,
$$
meeting the condition
\begin{equation}
\label{fd}
\psi = e^{i \langle k ,x \rangle} (1 + o(1)) \ \ \ \mbox{as $|x| \to \infty$}.
\end{equation}
Here $\langle x,y \rangle = \sum_{j=1}^n x_j y_j$ is the standard inner product in $\C^n$.

These functions were defined in \cite{Faddeev1965} and play the crucial role in
the multi-dimensional inverse scattering problem \cite{Faddeev,NK}.

In the sequel we consider two-dimensional Schr\"odinger operators:
$$
L = - \Delta + u = -\left(\frac{\partial^2}{\partial x_1^2} + \frac{\partial^2}{\partial x_2^2}\right) + u
= - 4\partial\bar{\partial} + u,
$$
where in the right-hand side we have
$\partial = \frac{\partial}{\partial z} =
\frac{1}{2}\left(\frac{\partial}{\partial x} - i \frac{\partial}{\partial y}\right),
\bar{\partial} = \frac{\partial}{\partial \bar{z}} =
\frac{1}{2}\left(\frac{\partial}{\partial x} + i \frac{\partial}{\partial y}\right),
(x_1,x_2) = (x,y) \in \R^2, z = x+iy \in \C$.

In this case, as well as for higher dimensions, there are not many examples for
which these eigenfunctions were computed explicitly. In fact, they are as follows:

1) for the multipoint potentials
$$
u(x) = \sum_{k=1}^N \varepsilon_k \delta(x-x_k),
$$
the Faddeev eigenfunctions are computed for all energy levels in \cite{GNR,GNR1};

2) for the Grinevich--Zakharov potentials \cite{G2} (these potentials
are reflectionless on this energy level $E$ and decay as $|x|^{-2}$)
explicit formulas for these Faddeev eigenfunctions on the energy level $E$
may be extracted from computations in \cite{G2}).

In \cite{GN} it is shown that for negative energy levels $E$
the functions
$$
\chi(x,k) = e^{-i\langle k,x\rangle}\psi(x,k) =
\exp \left(-\frac{i}{2}\sqrt{-E}\left(\lambda z - \frac{\bar{z}}{\lambda}\right)\right) \psi(x,k),
$$
where $k_1 = \frac{i\sqrt{-E}}{2}\left(\lambda +\frac{1}{\lambda}\right), k_2 =
-\frac{\sqrt{-E}}{2}\left(\lambda  - \frac{1}{\lambda}\right)$, satisfy the generalized
Cauchy--Riemann equation
$$
\frac{\partial \chi}{\partial \bar{\lambda}} = T \bar{\chi},
$$
i.e. they are generalized analytical functions, and if $T$ is nonsingular, then
this level is below the ground state. Moreover in the latter case
the inverse scattering
problem of reconstructing potentials (without assuming that these potentials
are sufficiently small) is solved by methods of generalized
analytic functions theory \cite{Vekua}.

In this article we consider the Faddeev eigenfunctions on the
zero energy level $E=0$. They meet the following condition
\begin{equation}
\label{asymptotic}
\psi(z,\bar{z},\lambda) = e^{\lambda z} \left(1 + \frac{A(\lambda,\bar{\lambda})}{z} +
e^{\bar{\lambda}\bar{z}-\lambda z}\frac{B(\lambda,\bar{\lambda})}{\bar{z}}
+ O\left(\frac{1}{|z|^2}\right)\right) 
\end{equation}
as $|z|\to \infty$.
In fact, this is one branch of these functions and another branch is
given by solutions to the equation $(-\Delta + u)\psi = 0$
with asymptotics $e^{\mu \bar{z}}(1 + o(1))$ as $|z|\to \infty$.
For non-zero energy levels the analogous asymptotic expansions were given
in \cite{Novikov1992} and for the zero energy level
it is derived in a similar way from the analytical properties of
the Green--Faddeev function of the Laplace operator.
The detailed analysis of the latter function, in particular,
was given in \cite{BLMP}.

\section{The Moutard transformation}

\subsection{The Moutard transformation and its double iteration}

Given $\omega$ satisfying the equation
$$
L \omega = (- \Delta + u )\omega = 0,
$$
{\it the Moutard transformation} of $L$
is defined as
$$
L \to \widetilde{L} = -\Delta + u -2\Delta \log \omega =
-\Delta - u + 2\frac{\omega_x^2+\omega_y^2}{\omega^2}.
$$
It is easy to check that

\begin{itemize}
\item
{\sl if $\varphi$ satisfies the equation $L \varphi = 0$,
then the function $\theta$ defined up to term $\frac{C}{\omega}, C = \mathrm{const}$,
from the system
\begin{equation}
\label{eigen-moutard}
(\omega \theta)_x = -\omega^2 \left(\frac{\varphi}{\omega}\right)_y, \ \ \
(\omega \theta)_y = \omega^2 \left(\frac{\varphi}{\omega}\right)_x
\end{equation}
satisfies}
$\widetilde{L}\theta = 0$.
\end{itemize}

If the potential $u$ depends only on $x$, then a certain reduction
of the Moutard transformation gives the famous Darboux transformation of one-dimensional
Schr\"odinger operators:
$$
L =  -\frac{d^2}{dx^2} + u = \left(\frac{d}{dx}+v\right)\left( - \frac{d}{dx}+v\right) \to
\widetilde{L} = \left(- \frac{d}{dx}+v\right)\left(\frac{d}{dx}+v\right).
$$
This is the case when
$\omega=f(x)e^{\sqrt{E}y}$ and  $\left( - \frac{d^2}{dx^2} + u\right) f = Ef$.

In \cite{TT} we used a double iteration of the Moutard
transformation starting from the Laplace operator $L_0 = -\Delta$.
Given two harmonic functions $\omega_1$ and $\omega_2$,
$\Delta\omega_1 = \Delta\omega_2 = 0$, we construct a pair of
Moutard transformations corresponding to $\omega_1$ and $\omega_2$:
$$
L_1 = -\Delta + u_1 \stackrel{\omega_1}{\longleftarrow} L_0
= -\Delta \stackrel{\omega_2}{\longrightarrow} L_2 = -\Delta + u_2.
$$
Both potentials $u_1=-2\Delta\log \omega_1$ and $u_2 = -2\Delta\log\omega_2$ have singularities at zeroes of
functions $\omega_1$ and $\omega_2$, respectively, and we have to iterate the Moutard transformation
to achieve, if possible,  smooth (i.e. bounded and differentiable) potentials.

\begin{figure}[htbp]
\begin{center}
 \unitlength 0.8mm \linethickness{0.4pt}
\begin{picture}(80.00,30.00)(0.0,75.0)
\multiput(9.00,91.00)(0.79,-0.09){2}{\line(1,0){0.79}}
\multiput(10.59,90.82)(0.30,-0.11){5}{\line(1,0){0.30}}
\multiput(12.09,90.28)(0.17,-0.11){8}{\line(1,0){0.17}}
\multiput(13.43,89.42)(0.11,-0.11){10}{\line(0,-1){0.11}}
\multiput(14.54,88.27)(0.12,-0.20){7}{\line(0,-1){0.20}}
\multiput(15.37,86.91)(0.10,-0.30){5}{\line(0,-1){0.30}}
\multiput(15.86,85.39)(0.07,-0.80){2}{\line(0,-1){0.80}}
\multiput(16.00,83.80)(-0.11,-0.79){2}{\line(0,-1){0.79}}
\multiput(15.77,82.22)(-0.12,-0.30){5}{\line(0,-1){0.30}}
\multiput(15.19,80.73)(-0.11,-0.16){8}{\line(0,-1){0.16}}
\multiput(14.29,79.42)(-0.13,-0.12){9}{\line(-1,0){0.13}}
\multiput(13.11,78.34)(-0.20,-0.11){7}{\line(-1,0){0.20}}
\multiput(11.72,77.55)(-0.38,-0.11){4}{\line(-1,0){0.38}}
\put(10.19,77.10){\line(-1,0){1.59}}
\multiput(8.60,77.01)(-0.52,0.09){3}{\line(-1,0){0.52}}
\multiput(7.03,77.28)(-0.25,0.10){6}{\line(-1,0){0.25}}
\multiput(5.56,77.90)(-0.16,0.12){8}{\line(-1,0){0.16}}
\multiput(4.27,78.84)(-0.12,0.13){9}{\line(0,1){0.13}}
\multiput(3.22,80.05)(-0.11,0.20){7}{\line(0,1){0.20}}
\multiput(2.48,81.46)(-0.10,0.39){4}{\line(0,1){0.39}}
\put(2.07,83.00){\line(0,1){1.60}}
\multiput(2.03,84.60)(0.11,0.52){3}{\line(0,1){0.52}}
\multiput(2.34,86.16)(0.11,0.24){6}{\line(0,1){0.24}}
\multiput(3.01,87.61)(0.11,0.14){9}{\line(0,1){0.14}}
\multiput(3.98,88.88)(0.14,0.11){9}{\line(1,0){0.14}}
\multiput(5.22,89.89)(0.24,0.12){6}{\line(1,0){0.24}}
\multiput(6.65,90.59)(0.59,0.10){4}{\line(1,0){0.59}}
\multiput(49.00,91.00)(0.79,-0.09){2}{\line(1,0){0.79}}
\multiput(50.59,90.82)(0.30,-0.11){5}{\line(1,0){0.30}}
\multiput(52.09,90.28)(0.17,-0.11){8}{\line(1,0){0.17}}
\multiput(53.43,89.42)(0.11,-0.11){10}{\line(0,-1){0.11}}
\multiput(54.54,88.27)(0.12,-0.20){7}{\line(0,-1){0.20}}
\multiput(55.37,86.91)(0.10,-0.30){5}{\line(0,-1){0.30}}
\multiput(55.86,85.39)(0.07,-0.80){2}{\line(0,-1){0.80}}
\multiput(56.00,83.80)(-0.11,-0.79){2}{\line(0,-1){0.79}}
\multiput(55.77,82.22)(-0.12,-0.30){5}{\line(0,-1){0.30}}
\multiput(55.19,80.73)(-0.11,-0.16){8}{\line(0,-1){0.16}}
\multiput(54.29,79.42)(-0.13,-0.12){9}{\line(-1,0){0.13}}
\multiput(53.11,78.34)(-0.20,-0.11){7}{\line(-1,0){0.20}}
\multiput(51.72,77.55)(-0.38,-0.11){4}{\line(-1,0){0.38}}
\put(50.19,77.10){\line(-1,0){1.59}}
\multiput(48.60,77.01)(-0.52,0.09){3}{\line(-1,0){0.52}}
\multiput(47.03,77.28)(-0.25,0.10){6}{\line(-1,0){0.25}}
\multiput(45.56,77.90)(-0.16,0.12){8}{\line(-1,0){0.16}}
\multiput(44.27,78.84)(-0.12,0.13){9}{\line(0,1){0.13}}
\multiput(43.22,80.05)(-0.11,0.20){7}{\line(0,1){0.20}}
\multiput(42.48,81.46)(-0.10,0.39){4}{\line(0,1){0.39}}
\put(42.07,83.00){\line(0,1){1.60}}
\multiput(42.03,84.60)(0.11,0.52){3}{\line(0,1){0.52}}
\multiput(42.34,86.16)(0.11,0.24){6}{\line(0,1){0.24}}
\multiput(43.01,87.61)(0.11,0.14){9}{\line(0,1){0.14}}
\multiput(43.98,88.88)(0.14,0.11){9}{\line(1,0){0.14}}
\multiput(45.22,89.89)(0.24,0.12){6}{\line(1,0){0.24}}
\multiput(46.65,90.59)(0.59,0.10){4}{\line(1,0){0.59}}
\multiput(29.00,106.00)(0.79,-0.09){2}{\line(1,0){0.79}}
\multiput(30.59,105.82)(0.30,-0.11){5}{\line(1,0){0.30}}
\multiput(32.09,105.28)(0.17,-0.11){8}{\line(1,0){0.17}}
\multiput(33.43,104.42)(0.11,-0.11){10}{\line(0,-1){0.11}}
\multiput(34.54,103.27)(0.12,-0.20){7}{\line(0,-1){0.20}}
\multiput(35.37,101.91)(0.10,-0.30){5}{\line(0,-1){0.30}}
\multiput(35.86,100.39)(0.07,-0.80){2}{\line(0,-1){0.80}}
\multiput(36.00,98.80)(-0.11,-0.79){2}{\line(0,-1){0.79}}
\multiput(35.77,97.22)(-0.12,-0.30){5}{\line(0,-1){0.30}}
\multiput(35.19,95.73)(-0.11,-0.16){8}{\line(0,-1){0.16}}
\multiput(34.29,94.42)(-0.13,-0.12){9}{\line(-1,0){0.13}}
\multiput(33.11,93.34)(-0.20,-0.11){7}{\line(-1,0){0.20}}
\multiput(31.72,92.55)(-0.38,-0.11){4}{\line(-1,0){0.38}}
\put(30.19,92.10){\line(-1,0){1.59}}
\multiput(28.60,92.01)(-0.52,0.09){3}{\line(-1,0){0.52}}
\multiput(27.03,92.28)(-0.25,0.10){6}{\line(-1,0){0.25}}
\multiput(25.56,92.90)(-0.16,0.12){8}{\line(-1,0){0.16}}
\multiput(24.27,93.84)(-0.12,0.13){9}{\line(0,1){0.13}}
\multiput(23.22,95.05)(-0.11,0.20){7}{\line(0,1){0.20}}
\multiput(22.48,96.46)(-0.10,0.39){4}{\line(0,1){0.39}}
\put(22.07,98.00){\line(0,1){1.60}}
\multiput(22.03,99.60)(0.11,0.52){3}{\line(0,1){0.52}}
\multiput(22.34,101.16)(0.11,0.24){6}{\line(0,1){0.24}}
\multiput(23.01,102.61)(0.11,0.14){9}{\line(0,1){0.14}}
\multiput(23.98,103.88)(0.14,0.11){9}{\line(1,0){0.14}}
\multiput(25.22,104.89)(0.24,0.12){6}{\line(1,0){0.24}}
\multiput(26.65,105.59)(0.59,0.10){4}{\line(1,0){0.59}}
\multiput(29.00,81.00)(0.79,-0.09){2}{\line(1,0){0.79}}
\multiput(30.59,80.82)(0.30,-0.11){5}{\line(1,0){0.30}}
\multiput(32.09,80.28)(0.17,-0.11){8}{\line(1,0){0.17}}
\multiput(33.43,79.42)(0.11,-0.11){10}{\line(0,-1){0.11}}
\multiput(34.54,78.27)(0.12,-0.20){7}{\line(0,-1){0.20}}
\multiput(35.37,76.91)(0.10,-0.30){5}{\line(0,-1){0.30}}
\multiput(35.86,75.39)(0.07,-0.80){2}{\line(0,-1){0.80}}
\multiput(36.00,73.80)(-0.11,-0.79){2}{\line(0,-1){0.79}}
\multiput(35.77,72.22)(-0.12,-0.30){5}{\line(0,-1){0.30}}
\multiput(35.19,70.73)(-0.11,-0.16){8}{\line(0,-1){0.16}}
\multiput(34.29,69.42)(-0.13,-0.12){9}{\line(-1,0){0.13}}
\multiput(33.11,68.34)(-0.20,-0.11){7}{\line(-1,0){0.20}}
\multiput(31.72,67.55)(-0.38,-0.11){4}{\line(-1,0){0.38}}
\put(30.19,67.10){\line(-1,0){1.59}}
\multiput(28.60,67.01)(-0.52,0.09){3}{\line(-1,0){0.52}}
\multiput(27.03,67.28)(-0.25,0.10){6}{\line(-1,0){0.25}}
\multiput(25.56,67.90)(-0.16,0.12){8}{\line(-1,0){0.16}}
\multiput(24.27,68.84)(-0.12,0.13){9}{\line(0,1){0.13}}
\multiput(23.22,70.05)(-0.11,0.20){7}{\line(0,1){0.20}}
\multiput(22.48,71.46)(-0.10,0.39){4}{\line(0,1){0.39}}
\put(22.07,73.00){\line(0,1){1.60}}
\multiput(22.03,74.60)(0.11,0.52){3}{\line(0,1){0.52}}
\multiput(22.34,76.16)(0.11,0.24){6}{\line(0,1){0.24}}
\multiput(23.01,77.61)(0.11,0.14){9}{\line(0,1){0.14}}
\multiput(23.98,78.88)(0.14,0.11){9}{\line(1,0){0.14}}
\multiput(25.22,79.89)(0.24,0.12){6}{\line(1,0){0.24}}
\multiput(26.65,80.59)(0.59,0.10){4}{\line(1,0){0.59}}
\put(9.00,84.00){\makebox(0,0)[cc]{$u_0$}}
\put(49.00,84.00){\makebox(0,0)[cc]{$u$}}
\put(29.00,99.00){\makebox(0,0)[cc]{$u_2$}}
\put(29.00,74.00){\makebox(0,0)[cc]{$u_1$}}
\put(42.00,73.39){\makebox(0,0)[cc]{$\theta_1$}}
\put(17.00,73.39){\makebox(0,0)[cc]{$\omega_1$}}
\put(16.33,95.79){\makebox(0,0)[cc]{$\omega_2$}}
\put(40.00,95.79){\makebox(0,0)[lc]{$\theta_2 = -
\frac{\omega_1\theta_1}{\omega_2}$}}
\put(29.00,44.00){\makebox(0,0)[cc]{~~~}}
\put(23.67,94.19){\line(-5,-3){9.67}}
\put(15.33,81.39){\line(2,-1){7.67}}
\put(34.67,77.39){\line(5,2){8.00}}
\put(42.67,87.79){\line(-4,3){8.33}}
\end{picture}
\end{center}
\caption{Rhombic superposition formula}
\label{fig-rhomb}
\end{figure}
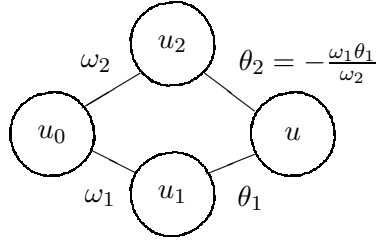

Let us denote by $\theta_1$ and $\theta_2$ the transforms of
$\omega_2$ and $\omega_1$ via (\ref{eigen-moutard}) for $\omega =
\omega_1$ and $\omega = \omega_2$, respectively. Such
transformations are defined up to $\frac{C}{\omega}$ and, for
$\theta_1$ fixed, we put $\theta_2 = -\frac{\omega_1}{\omega_2}
\theta_1$ (cf. Fig.~\ref{fig-rhomb}). We have
\begin{itemize}
\item
{\sl the functions $\theta_1$ and $\theta_2$ define the Moutard transformations of $L_2$ and $L_1$,
respectively, which give the same operator
$$
L_1 = -\Delta + u_1 \stackrel{\theta_1}{\longrightarrow} L = L_{12} = -\Delta + u _{12}
\stackrel{\theta_2}{\longleftarrow} L_2 = -\Delta + u_2.
$$

\item
the functions $\varphi_1 = \frac{1}{\theta_1}$ and $\varphi_2 =
\frac{1}{\theta_2}$ meet the equation $L \psi = 0$.}
\end{itemize}

The general formula for $u = u_{12}$ takes the form
\begin{equation}
\label{iteration}
u = - 2\Delta \log
i\left(
(p_1\bar{p}_2-p_2\bar{p}_1) + \right.
\end{equation}
$$
\left.
\int\left((p^\prime_1 p_2 - p_1 p^\prime_2)dz + (\bar{p}_1\bar{p}^\prime_2 -
\bar{p}^\prime_1 \bar{p}_2) d\bar{z}
\right)\right),
$$
where $\omega_1 = p_1(z)+\overline{p_1(z)}, \omega_2 = p_2(z) + \overline{p_2(z)}$, $p_1(z)$ and $p_2(z)$
are holomorphic functions of $z$, and the free scalar parameter which we mentioned above appears as
the integration constant in (\ref{iteration}).

\begin{figure}[htbp]
\begin{center}
 \unitlength 0.70mm \linethickness{0.4pt}
\begin{picture}(100.00,77.33)(70.0,70.0)
\multiput(102.00,147.33)(0.79,-0.09){2}{\line(1,0){0.79}}
\multiput(103.59,147.15)(0.30,-0.11){5}{\line(1,0){0.30}}
\multiput(105.09,146.61)(0.17,-0.11){8}{\line(1,0){0.17}}
\multiput(106.43,145.75)(0.11,-0.11){10}{\line(0,-1){0.11}}
\multiput(107.54,144.60)(0.12,-0.20){7}{\line(0,-1){0.20}}
\multiput(108.37,143.24)(0.10,-0.30){5}{\line(0,-1){0.30}}
\multiput(108.86,141.72)(0.07,-0.80){2}{\line(0,-1){0.80}}
\multiput(109.00,140.13)(-0.11,-0.79){2}{\line(0,-1){0.79}}
\multiput(108.77,138.55)(-0.12,-0.30){5}{\line(0,-1){0.30}}
\multiput(108.19,137.06)(-0.11,-0.16){8}{\line(0,-1){0.16}}
\multiput(107.29,135.75)(-0.13,-0.12){9}{\line(-1,0){0.13}}
\multiput(106.11,134.67)(-0.20,-0.11){7}{\line(-1,0){0.20}}
\multiput(104.72,133.88)(-0.38,-0.11){4}{\line(-1,0){0.38}}
\put(103.19,133.43){\line(-1,0){1.59}}
\multiput(101.60,133.34)(-0.52,0.09){3}{\line(-1,0){0.52}}
\multiput(100.03,133.61)(-0.25,0.10){6}{\line(-1,0){0.25}}
\multiput(98.56,134.23)(-0.16,0.12){8}{\line(-1,0){0.16}}
\multiput(97.27,135.17)(-0.12,0.13){9}{\line(0,1){0.13}}
\multiput(96.22,136.38)(-0.11,0.20){7}{\line(0,1){0.20}}
\multiput(95.48,137.79)(-0.10,0.39){4}{\line(0,1){0.39}}
\put(95.07,139.33){\line(0,1){1.60}}
\multiput(95.03,140.93)(0.11,0.52){3}{\line(0,1){0.52}}
\multiput(95.34,142.49)(0.11,0.24){6}{\line(0,1){0.24}}
\multiput(96.01,143.94)(0.11,0.14){9}{\line(0,1){0.14}}
\multiput(96.98,145.21)(0.14,0.11){9}{\line(1,0){0.14}}
\multiput(98.22,146.22)(0.24,0.12){6}{\line(1,0){0.24}}
\multiput(99.65,146.92)(0.59,0.10){4}{\line(1,0){0.59}}
\multiput(152.00,147.33)(0.79,-0.09){2}{\line(1,0){0.79}}
\multiput(153.59,147.15)(0.30,-0.11){5}{\line(1,0){0.30}}
\multiput(155.09,146.61)(0.17,-0.11){8}{\line(1,0){0.17}}
\multiput(156.43,145.75)(0.11,-0.11){10}{\line(0,-1){0.11}}
\multiput(157.54,144.60)(0.12,-0.20){7}{\line(0,-1){0.20}}
\multiput(158.37,143.24)(0.10,-0.30){5}{\line(0,-1){0.30}}
\multiput(158.86,141.72)(0.07,-0.80){2}{\line(0,-1){0.80}}
\multiput(159.00,140.13)(-0.11,-0.79){2}{\line(0,-1){0.79}}
\multiput(158.77,138.55)(-0.12,-0.30){5}{\line(0,-1){0.30}}
\multiput(158.19,137.06)(-0.11,-0.16){8}{\line(0,-1){0.16}}
\multiput(157.29,135.75)(-0.13,-0.12){9}{\line(-1,0){0.13}}
\multiput(156.11,134.67)(-0.20,-0.11){7}{\line(-1,0){0.20}}
\multiput(154.72,133.88)(-0.38,-0.11){4}{\line(-1,0){0.38}}
\put(153.19,133.43){\line(-1,0){1.59}}
\multiput(151.60,133.34)(-0.52,0.09){3}{\line(-1,0){0.52}}
\multiput(150.03,133.61)(-0.25,0.10){6}{\line(-1,0){0.25}}
\multiput(148.56,134.23)(-0.16,0.12){8}{\line(-1,0){0.16}}
\multiput(147.27,135.17)(-0.12,0.13){9}{\line(0,1){0.13}}
\multiput(146.22,136.38)(-0.11,0.20){7}{\line(0,1){0.20}}
\multiput(145.48,137.79)(-0.10,0.39){4}{\line(0,1){0.39}}
\put(145.07,139.33){\line(0,1){1.60}}
\multiput(145.03,140.93)(0.11,0.52){3}{\line(0,1){0.52}}
\multiput(145.34,142.49)(0.11,0.24){6}{\line(0,1){0.24}}
\multiput(146.01,143.94)(0.11,0.14){9}{\line(0,1){0.14}}
\multiput(146.98,145.21)(0.14,0.11){9}{\line(1,0){0.14}}
\multiput(148.22,146.22)(0.24,0.12){6}{\line(1,0){0.24}}
\multiput(149.65,146.92)(0.59,0.10){4}{\line(1,0){0.59}}
\multiput(132.00,127.33)(0.79,-0.09){2}{\line(1,0){0.79}}
\multiput(133.59,127.15)(0.30,-0.11){5}{\line(1,0){0.30}}
\multiput(135.09,126.61)(0.17,-0.11){8}{\line(1,0){0.17}}
\multiput(136.43,125.75)(0.11,-0.11){10}{\line(0,-1){0.11}}
\multiput(137.54,124.60)(0.12,-0.20){7}{\line(0,-1){0.20}}
\multiput(138.37,123.24)(0.10,-0.30){5}{\line(0,-1){0.30}}
\multiput(138.86,121.72)(0.07,-0.80){2}{\line(0,-1){0.80}}
\multiput(139.00,120.13)(-0.11,-0.79){2}{\line(0,-1){0.79}}
\multiput(138.77,118.55)(-0.12,-0.30){5}{\line(0,-1){0.30}}
\multiput(138.19,117.06)(-0.11,-0.16){8}{\line(0,-1){0.16}}
\multiput(137.29,115.75)(-0.13,-0.12){9}{\line(-1,0){0.13}}
\multiput(136.11,114.67)(-0.20,-0.11){7}{\line(-1,0){0.20}}
\multiput(134.72,113.88)(-0.38,-0.11){4}{\line(-1,0){0.38}}
\put(133.19,113.43){\line(-1,0){1.59}}
\multiput(131.60,113.34)(-0.52,0.09){3}{\line(-1,0){0.52}}
\multiput(130.03,113.61)(-0.25,0.10){6}{\line(-1,0){0.25}}
\multiput(128.56,114.23)(-0.16,0.12){8}{\line(-1,0){0.16}}
\multiput(127.27,115.17)(-0.12,0.13){9}{\line(0,1){0.13}}
\multiput(126.22,116.38)(-0.11,0.20){7}{\line(0,1){0.20}}
\multiput(125.48,117.79)(-0.10,0.39){4}{\line(0,1){0.39}}
\put(125.07,119.33){\line(0,1){1.60}}
\multiput(125.03,120.93)(0.11,0.52){3}{\line(0,1){0.52}}
\multiput(125.34,122.49)(0.11,0.24){6}{\line(0,1){0.24}}
\multiput(126.01,123.94)(0.11,0.14){9}{\line(0,1){0.14}}
\multiput(126.98,125.21)(0.14,0.11){9}{\line(1,0){0.14}}
\multiput(128.22,126.22)(0.24,0.12){6}{\line(1,0){0.24}}
\multiput(129.65,126.92)(0.59,0.10){4}{\line(1,0){0.59}}
\multiput(82.00,127.33)(0.79,-0.09){2}{\line(1,0){0.79}}
\multiput(83.59,127.15)(0.30,-0.11){5}{\line(1,0){0.30}}
\multiput(85.09,126.61)(0.17,-0.11){8}{\line(1,0){0.17}}
\multiput(86.43,125.75)(0.11,-0.11){10}{\line(0,-1){0.11}}
\multiput(87.54,124.60)(0.12,-0.20){7}{\line(0,-1){0.20}}
\multiput(88.37,123.24)(0.10,-0.30){5}{\line(0,-1){0.30}}
\multiput(88.86,121.72)(0.07,-0.80){2}{\line(0,-1){0.80}}
\multiput(89.00,120.13)(-0.11,-0.79){2}{\line(0,-1){0.79}}
\multiput(88.77,118.55)(-0.12,-0.30){5}{\line(0,-1){0.30}}
\multiput(88.19,117.06)(-0.11,-0.16){8}{\line(0,-1){0.16}}
\multiput(87.29,115.75)(-0.13,-0.12){9}{\line(-1,0){0.13}}
\multiput(86.11,114.67)(-0.20,-0.11){7}{\line(-1,0){0.20}}
\multiput(84.72,113.88)(-0.38,-0.11){4}{\line(-1,0){0.38}}
\put(83.19,113.43){\line(-1,0){1.59}}
\multiput(81.60,113.34)(-0.52,0.09){3}{\line(-1,0){0.52}}
\multiput(80.03,113.61)(-0.25,0.10){6}{\line(-1,0){0.25}}
\multiput(78.56,114.23)(-0.16,0.12){8}{\line(-1,0){0.16}}
\multiput(77.27,115.17)(-0.12,0.13){9}{\line(0,1){0.13}}
\multiput(76.22,116.38)(-0.11,0.20){7}{\line(0,1){0.20}}
\multiput(75.48,117.79)(-0.10,0.39){4}{\line(0,1){0.39}}
\put(75.07,119.33){\line(0,1){1.60}}
\multiput(75.03,120.93)(0.11,0.52){3}{\line(0,1){0.52}}
\multiput(75.34,122.49)(0.11,0.24){6}{\line(0,1){0.24}}
\multiput(76.01,123.94)(0.11,0.14){9}{\line(0,1){0.14}}
\multiput(76.98,125.21)(0.14,0.11){9}{\line(1,0){0.14}}
\multiput(78.22,126.22)(0.24,0.12){6}{\line(1,0){0.24}}
\multiput(79.65,126.92)(0.59,0.10){4}{\line(1,0){0.59}}
\multiput(82.00,77.33)(0.79,-0.09){2}{\line(1,0){0.79}}
\multiput(83.59,77.15)(0.30,-0.11){5}{\line(1,0){0.30}}
\multiput(85.09,76.61)(0.17,-0.11){8}{\line(1,0){0.17}}
\multiput(86.43,75.75)(0.11,-0.11){10}{\line(0,-1){0.11}}
\multiput(87.54,74.60)(0.12,-0.20){7}{\line(0,-1){0.20}}
\multiput(88.37,73.24)(0.10,-0.30){5}{\line(0,-1){0.30}}
\multiput(88.86,71.72)(0.07,-0.80){2}{\line(0,-1){0.80}}
\multiput(89.00,70.13)(-0.11,-0.79){2}{\line(0,-1){0.79}}
\multiput(88.77,68.55)(-0.12,-0.30){5}{\line(0,-1){0.30}}
\multiput(88.19,67.06)(-0.11,-0.16){8}{\line(0,-1){0.16}}
\multiput(87.29,65.75)(-0.13,-0.12){9}{\line(-1,0){0.13}}
\multiput(86.11,64.67)(-0.20,-0.11){7}{\line(-1,0){0.20}}
\multiput(84.72,63.88)(-0.38,-0.11){4}{\line(-1,0){0.38}}
\put(83.19,63.43){\line(-1,0){1.59}}
\multiput(81.60,63.34)(-0.52,0.09){3}{\line(-1,0){0.52}}
\multiput(80.03,63.61)(-0.25,0.10){6}{\line(-1,0){0.25}}
\multiput(78.56,64.23)(-0.16,0.12){8}{\line(-1,0){0.16}}
\multiput(77.27,65.17)(-0.12,0.13){9}{\line(0,1){0.13}}
\multiput(76.22,66.38)(-0.11,0.20){7}{\line(0,1){0.20}}
\multiput(75.48,67.79)(-0.10,0.39){4}{\line(0,1){0.39}}
\put(75.07,69.33){\line(0,1){1.60}}
\multiput(75.03,70.93)(0.11,0.52){3}{\line(0,1){0.52}}
\multiput(75.34,72.49)(0.11,0.24){6}{\line(0,1){0.24}}
\multiput(76.01,73.94)(0.11,0.14){9}{\line(0,1){0.14}}
\multiput(76.98,75.21)(0.14,0.11){9}{\line(1,0){0.14}}
\multiput(78.22,76.22)(0.24,0.12){6}{\line(1,0){0.24}}
\multiput(79.65,76.92)(0.59,0.10){4}{\line(1,0){0.59}}
\multiput(132.00,77.33)(0.79,-0.09){2}{\line(1,0){0.79}}
\multiput(133.59,77.15)(0.30,-0.11){5}{\line(1,0){0.30}}
\multiput(135.09,76.61)(0.17,-0.11){8}{\line(1,0){0.17}}
\multiput(136.43,75.75)(0.11,-0.11){10}{\line(0,-1){0.11}}
\multiput(137.54,74.60)(0.12,-0.20){7}{\line(0,-1){0.20}}
\multiput(138.37,73.24)(0.10,-0.30){5}{\line(0,-1){0.30}}
\multiput(138.86,71.72)(0.07,-0.80){2}{\line(0,-1){0.80}}
\multiput(139.00,70.13)(-0.11,-0.79){2}{\line(0,-1){0.79}}
\multiput(138.77,68.55)(-0.12,-0.30){5}{\line(0,-1){0.30}}
\multiput(138.19,67.06)(-0.11,-0.16){8}{\line(0,-1){0.16}}
\multiput(137.29,65.75)(-0.13,-0.12){9}{\line(-1,0){0.13}}
\multiput(136.11,64.67)(-0.20,-0.11){7}{\line(-1,0){0.20}}
\multiput(134.72,63.88)(-0.38,-0.11){4}{\line(-1,0){0.38}}
\put(133.19,63.43){\line(-1,0){1.59}}
\multiput(131.60,63.34)(-0.52,0.09){3}{\line(-1,0){0.52}}
\multiput(130.03,63.61)(-0.25,0.10){6}{\line(-1,0){0.25}}
\multiput(128.56,64.23)(-0.16,0.12){8}{\line(-1,0){0.16}}
\multiput(127.27,65.17)(-0.12,0.13){9}{\line(0,1){0.13}}
\multiput(126.22,66.38)(-0.11,0.20){7}{\line(0,1){0.20}}
\multiput(125.48,67.79)(-0.10,0.39){4}{\line(0,1){0.39}}
\put(125.07,69.33){\line(0,1){1.60}}
\multiput(125.03,70.93)(0.11,0.52){3}{\line(0,1){0.52}}
\multiput(125.34,72.49)(0.11,0.24){6}{\line(0,1){0.24}}
\multiput(126.01,73.94)(0.11,0.14){9}{\line(0,1){0.14}}
\multiput(126.98,75.21)(0.14,0.11){9}{\line(1,0){0.14}}
\multiput(128.22,76.22)(0.24,0.12){6}{\line(1,0){0.24}}
\multiput(129.65,76.92)(0.59,0.10){4}{\line(1,0){0.59}}
\multiput(102.00,97.33)(0.79,-0.09){2}{\line(1,0){0.79}}
\multiput(103.59,97.15)(0.30,-0.11){5}{\line(1,0){0.30}}
\multiput(105.09,96.61)(0.17,-0.11){8}{\line(1,0){0.17}}
\multiput(106.43,95.75)(0.11,-0.11){10}{\line(0,-1){0.11}}
\multiput(107.54,94.60)(0.12,-0.20){7}{\line(0,-1){0.20}}
\multiput(108.37,93.24)(0.10,-0.30){5}{\line(0,-1){0.30}}
\multiput(108.86,91.72)(0.07,-0.80){2}{\line(0,-1){0.80}}
\multiput(109.00,90.13)(-0.11,-0.79){2}{\line(0,-1){0.79}}
\multiput(108.77,88.55)(-0.12,-0.30){5}{\line(0,-1){0.30}}
\multiput(108.19,87.06)(-0.11,-0.16){8}{\line(0,-1){0.16}}
\multiput(107.29,85.75)(-0.13,-0.12){9}{\line(-1,0){0.13}}
\multiput(106.11,84.67)(-0.20,-0.11){7}{\line(-1,0){0.20}}
\multiput(104.72,83.88)(-0.38,-0.11){4}{\line(-1,0){0.38}}
\put(103.19,83.43){\line(-1,0){1.59}}
\multiput(101.60,83.34)(-0.52,0.09){3}{\line(-1,0){0.52}}
\multiput(100.03,83.61)(-0.25,0.10){6}{\line(-1,0){0.25}}
\multiput(98.56,84.23)(-0.16,0.12){8}{\line(-1,0){0.16}}
\multiput(97.27,85.17)(-0.12,0.13){9}{\line(0,1){0.13}}
\multiput(96.22,86.38)(-0.11,0.20){7}{\line(0,1){0.20}}
\multiput(95.48,87.79)(-0.10,0.39){4}{\line(0,1){0.39}}
\put(95.07,89.33){\line(0,1){1.60}}
\multiput(95.03,90.93)(0.11,0.52){3}{\line(0,1){0.52}}
\multiput(95.34,92.49)(0.11,0.24){6}{\line(0,1){0.24}}
\multiput(96.01,93.94)(0.11,0.14){9}{\line(0,1){0.14}}
\multiput(96.98,95.21)(0.14,0.11){9}{\line(1,0){0.14}}
\multiput(98.22,96.22)(0.24,0.12){6}{\line(1,0){0.24}}
\multiput(99.65,96.92)(0.59,0.10){4}{\line(1,0){0.59}}
\multiput(152.00,97.33)(0.79,-0.09){2}{\line(1,0){0.79}}
\multiput(153.59,97.15)(0.30,-0.11){5}{\line(1,0){0.30}}
\multiput(155.09,96.61)(0.17,-0.11){8}{\line(1,0){0.17}}
\multiput(156.43,95.75)(0.11,-0.11){10}{\line(0,-1){0.11}}
\multiput(157.54,94.60)(0.12,-0.20){7}{\line(0,-1){0.20}}
\multiput(158.37,93.24)(0.10,-0.30){5}{\line(0,-1){0.30}}
\multiput(158.86,91.72)(0.07,-0.80){2}{\line(0,-1){0.80}}
\multiput(159.00,90.13)(-0.11,-0.79){2}{\line(0,-1){0.79}}
\multiput(158.77,88.55)(-0.12,-0.30){5}{\line(0,-1){0.30}}
\multiput(158.19,87.06)(-0.11,-0.16){8}{\line(0,-1){0.16}}
\multiput(157.29,85.75)(-0.13,-0.12){9}{\line(-1,0){0.13}}
\multiput(156.11,84.67)(-0.20,-0.11){7}{\line(-1,0){0.20}}
\multiput(154.72,83.88)(-0.38,-0.11){4}{\line(-1,0){0.38}}
\put(153.19,83.43){\line(-1,0){1.59}}
\multiput(151.60,83.34)(-0.52,0.09){3}{\line(-1,0){0.52}}
\multiput(150.03,83.61)(-0.25,0.10){6}{\line(-1,0){0.25}}
\multiput(148.56,84.23)(-0.16,0.12){8}{\line(-1,0){0.16}}
\multiput(147.27,85.17)(-0.12,0.13){9}{\line(0,1){0.13}}
\multiput(146.22,86.38)(-0.11,0.20){7}{\line(0,1){0.20}}
\multiput(145.48,87.79)(-0.10,0.39){4}{\line(0,1){0.39}}
\put(145.07,89.33){\line(0,1){1.60}}
\multiput(145.03,90.93)(0.11,0.52){3}{\line(0,1){0.52}}
\multiput(145.34,92.49)(0.11,0.24){6}{\line(0,1){0.24}}
\multiput(146.01,93.94)(0.11,0.14){9}{\line(0,1){0.14}}
\multiput(146.98,95.21)(0.14,0.11){9}{\line(1,0){0.14}}
\multiput(148.22,96.22)(0.24,0.12){6}{\line(1,0){0.24}}
\multiput(149.65,96.92)(0.59,0.10){4}{\line(1,0){0.59}}
\put(82.00,70.33){\makebox(0,0)[cc]{$u_0$}}
\put(132.00,70.33){\makebox(0,0)[cc]{$u_1$}}
\put(102.00,90.33){\makebox(0,0)[cc]{$u_2$}}
\put(152.00,90.33){\makebox(0,0)[cc]{$u$}}
\put(146.67,77.12){\makebox(0,0)[cc]{$\theta_1$}}
\put(123.00,97.12){\makebox(0,0)[cc]{$\theta_2$}}
\put(89.33,87.52){\makebox(0,0)[cc]{$\omega _2$}}
\put(107.00,63.52){\makebox(0,0)[cc]{$\omega _1$}}
\put(76.33,95.52){\makebox(0,0)[cc]{$\psi_0$}}
\put(97.00,110.33){\makebox(0,0)[cc]{$\psi_2$}}
\put(136.67,101.12){\makebox(0,0)[cc]{$\psi_1$}} 
\put(158.00,115.52){\makebox(0,0)[lc]{$\psi$}}
\put(117.00,45.33){\makebox(0,0)[cc]{~~~}}
\put(29.00,44.00){\makebox(0,0)[cc]{~~~}}
\put(82.00,113.92){\line(0,-1){36.80}}
\put(89.00,70.72){\line(1,0){36.00}}
\put(145.00,90.72){\line(-1,0){35.67}}
\put(102.00,97.12){\line(0,1){36.80}}
\put(109.00,140.32){\line(1,0){36.00}}
\put(152.00,133.92){\line(0,-1){36.80}}
\put(132.00,77.12){\line(0,1){36.80}}
\put(125.00,120.32){\line(-1,0){36.00}}
\put(147.00,135.33){\line(-1,-1){10.67}}
\put(137.33,75.52){\line(1,1){10.33}}
\put(97.33,85.12){\line(-6,-5){10.67}}
\put(97.33,134.72){\line(-5,-4){11.00}}
\end{picture}
\end{center}
\caption{Cubic superposition formula}
\label{fig-cube}
\end{figure}
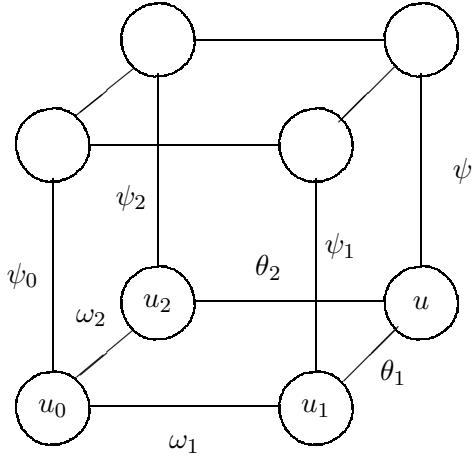

\subsection{The Moutard transformation of Faddeev eigenfunctions}

For $L_0 = -\Delta$ the Faddeev eigenfunctions (\ref{asymptotic})
are
$$
\psi_0(z,\lambda) = e^{\lambda z}
$$
and
they are transformed via (\ref{eigen-moutard}) to $\psi_1$ and $\psi_2$:
$$
\psi_1  \stackrel{\omega_1}{\longleftarrow} \psi_0 = e^{\lambda z}
\stackrel{\omega_2}{\longrightarrow} \psi_2.
$$
As we mentioned above these transformations are not uniquely defined
however at least for functions $\omega_1$ and $\omega_2$ which are
polynomial in $z$ and $\bar{z}$ it is possible to choose the
branches of $\psi_1$ and $\psi_2$ such that they would meet the
condition (\ref{fd}). Then ``the cubic superposition formula'' (see
\cite{TT} and Fig.~\ref{fig-cube}) implies that
$$
\psi(z,\bar{z},\lambda) = e^{\lambda z} +
\frac{\omega_2}{\theta_1}(\psi_2 - \psi_1)
$$
satisfies  the equation $L \psi = 0$. It
seems that for smooth fast decaying potentials $u=u_{12}$ the
function $\psi$ is the Faddeev eigenfunction on the zero energy
level. We do not state that here as a rigorous mathematical result
however for the potentials found in \cite{TT} this is the case.

{\sc Example}.
Let
$$
p_1(z) = \left(1 - \frac{i}{4}\right) z^2 +
\frac{z}{2}, \ \ \ p_2(z)  = \frac{1}{4}(3-5i)z^2 + \frac{1-i}{2}z.
$$
For some appropriate  constant  $C$ in $\theta_1$ we obtain
$$
u =  -\frac{5120|(4-i)z+1|^2}{(160+|z|^2|(4-i)z+2|^2)^2}.
$$
After simplifying
multiplications by some constants
$\varphi_1$ and $\varphi_2$ take the form
$$
\varphi_1 = \frac{2(z+\bar{z}) + (4-i)z^2 + (4+i)\bar{z}^2}{160+|z|^2|(4-i)z+2|^2},
$$
$$
\varphi_2 = \frac{2(1-i)z + 2(1+i) \bar{z}+ (3-5i)z^2 + (3+5i)\bar{z}^2}{160+|z|^2|(4-i)z+2|^2}.
$$
The Faddeev eigenfunctions (\ref{asymptotic}) for $L = -\Delta+u$ are as follows:
$$
\psi(z,\bar{z},\lambda) = e^{\lambda z} \left(1 +
\frac{1}{\lambda} \frac{(8i -32) z\bar{z}- 8 \bar{z} - (16+4i) \bar{z}^2 - 68 z\bar{z}^2}
{160+|z|^2|(4-i)z+2|^2} + \right.
$$
$$
\left.
+\frac{1}{\lambda^2}\frac{(32-8i) \bar{z} + 68 \bar{z}^2}{160+|z|^2|(4-i)z+2|^2}\right) =
$$
$$
= e^{\lambda z}\left( 1 - \frac{4}{\lambda z} + O\left(\frac{1}{|z|^2}\right)\right) \ \ \
\mbox{as $|z| \to \infty$}.
$$
The functions $u, \varphi_1$, and $\varphi_2$ are smooth and
real-valued, and the function $\psi$ is also smooth in $z,\bar{z}$.
The potential $u$ decays like $1/|z|^6$ as $|z|
\to \infty$, and $\varphi_1$ and $\varphi_2$ decay like $1/|z|^2$
as $|z| \to \infty$. Hence $\varphi_1$ and $\varphi_2$ span a
two-dimensional subspace in the kernel of $L$ and, since $u \leq 0$
and $u$ decays sufficiently fast, there are also negative eigenvalues of $L$.

We also see that ``the scattering data'' takes a very simple form:
$$
A(\lambda,\bar{\lambda}) = -\frac{4}{\lambda}, \ \ \ B(\lambda,\bar{\lambda}) = 0.
$$

Similar formulas for the Faddeev eigenfunctions may be easily derived
for another smooth and fast decaying potential
with non-trivial kernel which was found in \cite{TT} and corresponds to
$p_1(z)\! = \!(i\!-\! 1)z^3\! + \!\left(\frac{1}{10}\! +\!
\frac{3}{20}i\!\right)z^2 \!\!+\! \frac{z}{2}, \
 p_2(z)\! =\! 2iz^3\!\! +\!
\left(\frac{1}{4}\!+\!\frac{i}{20}\right)z^2\!\! +\!
\frac{1\!-\!i}{2}z$.
In this case ``the scattering data'' are
$A(\lambda,\bar{\lambda})=-\frac{6}{\lambda}$,  $B(\lambda,\bar{\lambda}) =
0$.

\section{The extended Moutard transformation and the Novikov--Veselov equation}

\subsection{The extended Moutard transformation}

In order to comply with the traditional notation related to
the Novikov--Veselov equation we renormalize the Schr\"odinger
operator as follows
$$
H = \partial\bar{\partial} + U = \frac{1}{4}\Delta - \frac{u}{4}.
$$
The Moutard transformation takes the form
$$
U \to U + 2 \partial\bar{\partial} \log \omega,
$$
$$
\left(\bar{\partial} + \frac{\omega_{\bar{z}}}{\omega}\right) \theta =
i \left(\bar{\partial} - \frac{\omega_{\bar{z}}}{\omega}\right) \varphi,
\ \ \
\left(\partial + \frac{\omega_z}{\omega}\right) \theta =
-i \left(\partial - \frac{\omega_z}{\omega}\right) \varphi.
$$

Let us assume that $\varphi$ depends also on the temporal variable $t$
and satisfies the equations
\begin{equation}
\label{eq}
H \varphi = 0, \ \ \
\partial_t \varphi = (\partial^3 + \bar{\partial}^3 + 3V\partial +
3\bar{V}\bar{\partial})\varphi, \ \ \
\bar{\partial}V = \partial U
\end{equation}
where $U = \bar{U}$. It is checked by straightforward computations that

\begin{itemize}
\item
{\sl given real-valued functions $\omega$ and $U$, the system (\ref{eq}) is invariant under the
extended Moutard transformation \cite{TT,HLL}}
\begin{equation}
\label{ext}
\begin{split}
\varphi \to \theta = \frac{i}{\omega} \int (\varphi\partial \omega -
\omega \partial \varphi)dz - (\varphi \bar{\partial} \omega - \omega
\bar{\partial}\varphi)d\bar{z} +
\\
[\varphi\partial^3\omega - \omega\partial^3\varphi +
\omega\bar{\partial}^3\varphi - \varphi\bar{\partial}^3\omega
+2(\partial^2\varphi\partial\omega - \partial\varphi\partial^2\omega) -
\\
 2(\bar{\partial}^2\varphi\bar{\partial}\omega -
\bar{\partial}\varphi\bar{\partial}^2\omega)
+ 3V(\varphi\partial\omega  - \omega\partial\varphi)
+ 3\bar{V}(\omega\bar{\partial}\varphi   - \varphi\bar{\partial}\omega)]dt,
\\
U \to U + 2\partial\bar{\partial}\log \omega, \ \
V \to V + 2 \partial^2 \log \omega.
\end{split}
\end{equation}
\end{itemize}

The compatibility condition for (\ref{eq}) is
the well-known Novikov--Veselov equation \cite{VN}
\begin{equation}
\label{nv}
U_t = \partial^3 U + \bar{\partial}^3 U + 3\partial(VU) +
3\bar{\partial} (\bar{V} U) =0, \ \ \
\bar{\partial}V = \partial U.
\end{equation}
It is a two-dimensional generalization of the
Korteweg--de Vries equation to which it reduces in the case $U=U(x), U=V$.

We may obtain analogs of all constructions of \S 2 by replacing the Moutard transformation
by its extended version (\ref{ext}) and construct by this mean non-trivial real-valued
solutions of the Novikov--Veselov equation (\ref{nv})
(cf. \cite{TT}). The analog of (\ref{iteration}) takes the form
$$
U(z,\bar{z},t) = 2\partial\bar{\partial} \log i \Big(
(p_1\bar{p_2}-p_2\bar{p_1}) + \int\left((p_1^\prime p_2 - p_1
p_2^\prime)dz +
\right.
$$
\begin{equation}
\label{extiteration}
\left.
(\bar{p_1}\bar{p_2}^\prime - \bar{p_1}^\prime
\bar{p_2}) d\bar{z}\right)
+ \int (p_1^{\prime\prime\prime}p_2 - p_1p_2^{\prime\prime\prime} +
2(p_1^\prime p_2^{\prime\prime} - p_1^{\prime\prime}p_2^\prime) +
\end{equation}
$$
+
\bar{p_1}\bar{p_2}^{\prime\prime\prime} -
\bar{p_1}^{\prime\prime\prime}\bar{p_2} +
2(\bar{p_1}^{\prime\prime}\bar{p_2}^\prime -
\bar{p_1}^\prime\bar{p_2}^{\prime\prime}))dt \Big)
$$
and, if the holomorphic in $z$ functions $p_1(z,t)$ and $p_2(z,t)$
satisfy the equation
$$
\frac{\partial p}{\partial t} = \frac{\partial^3 p}{\partial^3 z},
$$
then $U(z,\bar{z},t)$ satisfies the Novikov--Veselov equation (\ref{nv}).

\subsection{An example of the NV evolution of Faddeev eigenfunctions}

Let
$$
p_1(z) =iz^2, \ \ \
p_2=(1+i)z+z^2.
$$
Although these functions are independent on
$t$ the formula (\ref{extiteration}) gives a non-stationary solution
of the Novikov--Veselov equation. The Faddeev eigenfunctions are obtained by applying
iterations of (\ref{ext}) to $\psi_0(z,\bar{z},t) = e^{\lambda z + \lambda^3 t}$.

These data lead to the following functions:
$$
U(z,\bar{z},t) =
\frac{F_U(z,\bar{z},t)}{Q^2(z,\bar{z},t)}, \ \ \
V(z,\bar{z},t) = \frac{F_V(z,\bar{z},t)}{Q^2(z,\bar{z},t)}
$$
where
$$
F_U(z,\bar{z},T)  = 6i ( 24t (i\bar{z} - iz + z + \bar{z}) + 2 i z^4 \bar{z} + 4 i z^3 \bar{z} - 2 i z \bar{z}^4
 - 4 i z \bar{z}^3 +
$$
$$
+
60 i z - 60 i \bar{z} + 96 t z \bar{z} +
2 z^4 \bar{z} + z^4 + 6 z^2 \bar{z}^2 + 2 z \bar{z}^4 - 240 z \bar{z} - 60 z + \bar{z}^4
 - 60 \bar{z}),
$$
$$
F_V (z,\bar{z},t) = 6i (24t( iz - i\bar{z} + z + \bar{z}) + i z^4 + 4 i z^3 \bar{z}^2 - 12 i z^2 \bar{z}^3
 - 6 i z^2 \bar{z}^2 +
$$
$$
+6 i z \bar{z}^4 - 60 i z + 2 i \bar{z}^5 +
5 i \bar{z}^4 + 60 i \bar{z}
 + 48 t \bar{z}^2 + 4 z^3 \bar{z}^2 + 4 z^3 \bar{z} + 12 z^2 \bar{z}^4
 + 12 z^2 \bar{z}^3 +
$$
$$
+6 z \bar{z}^4 + 4 z \bar{z}^3 - 60 z - 2 \bar{z}^5 - 120 \bar{z}^2 - 60 \bar{z}),
$$
and
$$
Q(z,\bar{z},t) = (12 - 12 i) t - z^3 - (3 - 3 i) z^2 \bar{z}^2 + 3 i z^2 \bar{z} - 3 z \bar{z}^2
+ i \bar{z}^3 - (30 - 30 i) =
$$
$$
=\frac{1}{1+i}\left(24 t - \left[6(x^2+y^2)^2 + 8 (x^3 + y^3) + 60\right]\right).
$$
The Faddeev eigenfunctions for the potentials $U(z,\bar{z},t)$ have the form
$$
\psi(z,\bar{z},t,\lambda) =
e^{\lambda z}\left(1+ \frac{\mu_1(z,\bar{z},t)}{\lambda} + \frac{\mu_2(z,\bar{z},t)}{\lambda^2}\right) =
$$
$$
 = e^{\lambda z}
\Big( 1
+
\frac{1}{\lambda}
\frac{6( - 2 i z \bar{z}^2 - 2 i z \bar{z} + z^2 + 2 z \bar{z}^2 + \bar{z}^2)}
{Q(z,\bar{z},t)} +
$$
$$
+\frac{1}{\lambda^2} \frac{12 (i \bar{z}^2 + i \bar{z} - z - \bar{z}^2)}
{Q(z,\bar{z},t)}\Big).
$$
From the latter formula it is easy to derive ``the scattering data'' for $U(z,\bar{z},t)$:
$$
\psi(z,\bar{z},t,\lambda) = e^{\lambda z} \left(1 - \frac{4}{\lambda z} + O\left(\frac{1}{|z|^2}\right)\right)
\ \ \ \mbox{as $|z| \to \infty$},
$$
i.e. we have
$$
A(\lambda,\bar{\lambda},t) = -\frac{4}{\lambda}, \ \ \ B(\lambda,\bar{\lambda},t) = 0.
$$

The function $U(z,\bar{z},t)$ is a solution to the Novi\-kov--Veselov equation
with smooth and fast decaying (like $1/|z|^3$ as $|z| \to \infty$)
initial data $U(z,\bar{z},0)$ at $t=0$ that blows up at finite time $t = T_\ast$ \cite{TT}.
The blow up happens when $Q(z,\bar{z},t)$ vanishes at some point.

We note that

\begin{itemize}

\item
{\sl ``the scattering data'' $A(\lambda,\bar{\lambda},t)$ and $B(\lambda,\bar{\lambda},t)$ are
stationary;}

\item
{\sl the real and imaginary parts of $\mu_2(z,\bar{z},t)$
give eigenfunctions of $H$ at the zero
energy level. By the formula for $\mu_2(z,\bar{z},t)$, they
stay nonsingular and quadratically integrable until the
critical time $t=T_\ast$ when they achieve the same singularities as $U$.}

\end{itemize}

\section{Final remarks} 

We think that these examples are helpful in understanding the scattering transform for
the two-dimensional Schr\"odinger equation at zero energy and the regularity of the Cauchy problem for the Novikov--Veselov (NV) equation.  Therewith we have to mention a couple of results which were very recently obtained in this direction:

1) the scheme of inverse scattering for the NV equation was realized for initial data of conductivity type
\cite{P}. We recall that a conductivity type potential is just a potential of the form $u=\frac{\Delta \psi}{\psi}$ with  $\psi >0$ everywhere and 
$\psi =1$ near infinity and that operators with conductivity type potentials have no $L_2$-eigenvalues;

2) In \cite{GN} the first example of a potential with a nontrivial
exceptional set E, i.e. set consisting of $\lambda \in \C$ for which
the Faddeev eigenfunction (\ref{asymptotic}) is not uniquely defined,
was found. That was established for a point
potential $u(x) = \varepsilon \delta(x)$. Recently that study was
extended in \cite{MPS} where the first smooth examples of such potentials
were found and also a dichotomy type conjecture
on the regularity of the Cauchy problem for
the NV equation was introduced.


\vskip2mm {\sc Acknowledgement.} The work was supported by the
interdisciplinary project ``Geometrical and algebraic methods of
finding explicit solutions to equations of mathematical physics and
continuum mechanics'' of SB RAS and by the program ``Scientific Schools'' (grant 544.2012.1).
The authors thank P.G. Grinevich and R.G.
Novikov for helpful discussions.

\newpage

\end{document}